\documentclass[prd,amsmath,preprintnumbers,superscriptaddress,floatfix,letterpaper,showpacs,english]{revtex4}
\usepackage{amssymb}
\usepackage{amsmath}
\usepackage{amsfonts}
\usepackage{epsfig}
\usepackage{graphicx}
\usepackage{tabularx}
\usepackage{latexsym}
\usepackage{subfigure}
\usepackage{verbatim}
\usepackage{color}
\usepackage{braket}
\usepackage{multirow}
\usepackage{dcolumn}% Align table columns on decimal point
\usepackage{mathtools}
\usepackage{placeins}
\usepackage{epstopdf}
\usepackage{wrapfig}
\usepackage{hyperref}
\usepackage{diagbox}
\usepackage{booktabs}
\usepackage{simplewick}
\usepackage[normalem]{ulem}
\usepackage[latin1]{inputenc}
\usepackage{extarrows}
\usepackage{lipsum,babel}
\usepackage[toc,page]{appendix}

\def\gev{\mathrm{GeV}}

\def\lsim{\raise0.3ex\hbox{$<$\kern-0.75em\raise-1.1ex\hbox{$\sim$}}}
\def\gsim{\raise0.3ex\hbox{$>$\kern-0.75em\raise-1.1ex\hbox{$\sim$}}}

\newcommand{\bee}{\begin{equation}}
\newcommand{\ee}{\end{equation}}

\renewcommand\sout{\bgroup \color{red} \ULdepth=-.5ex \ULset}

\newcommand{\half}{ {\textstyle\frac{1}{2}} }
\renewcommand{\slash}[1]{#1 \hspace{-0.45em} / }
\usepackage{mathtools}
\DeclarePairedDelimiter\abs{\lvert}{\rvert}%
\allowdisplaybreaks[1] % n=0,1,2,3,4

\def\eq{\begin{eqnarray}}
\def\en{\end{eqnarray}}

\begin{document}
\preprint{}

%Title of paper
\title{Polarized GPDs and structure functions of $\rho$ meson}

\author{Bao-Dong Sun}
\email{sunbd@ihep.ac.cn}
\affiliation{Institute of High Energy Physics, Chinese Academy of Sciences, Beijing 100049, People's Republic of China}
\affiliation{School of Physics, University of Chinese Academy of Sciences, Beijing 100049, People's Republic of China}
\author{Yu-Bing Dong}
\affiliation{Institute of High Energy Physics, Chinese Academy of Sciences, Beijing 100049, People's Republic of China}
\affiliation{School of Physics,  University of Chinese Academy of Sciences, Beijing 100049, People's Republic of China}
\affiliation{Theoretical Physics Center for Science Facilities (TPCSF), CAS, Beijing 100049, People's Republic of China}
\noaffiliation

\begin{abstract}
The $\rho$ meson polarized generalized parton distribution functions, its structure functions $g_1$ and
$g_2$  and its axial form factors ${\tilde G}_{1,2}$ are studied based on a light-front quark model for the first
time.  Comparing our obtained moments of $g_1$ to the Lattice QCD calculation, we find that our results are
reasonably consistent to the Lattice predictions.
\end{abstract}
\date{\today}
\maketitle

\section{Introduction}
\label{intro}
It is believed that the generalized parton distributions (GPDs) of a system  could be a powerful tool to understand
its hadronic structure~\cite{Diehl:2003ny}. This is because GPDs naturally embody the information of both form
factors (FFs) and parton distribution functions (PDFs) for the complicated system. They can provide the normal
PDFs for the longitudinal parton distribution as well as the transverse information. Consequentially, GPDs
display the unique properties to present a "three-dimensional (3D)" description for the transverse and
longitudinal partonic degrees of freedom inside the system. Furthermore, it should be addressed that the
physical meaning of the transverse distribution is more transparent when one goes to the impact parameter
space~\cite{Burkardt:2002hr,Miller:2010nz,Sun:2018tmk}. Another important potential of GPDs is the information
about how the orbital angular momentum contributes to the total spin of a hadron. We know that the sum rules
proposed by Xiangdong Ji for a nucleon (spin-1/2) reveal the relation between GPDs and the spin carried by
quarks and gluons~\cite{Ji:1996nm,Ji:1998pc}. For the spin-1 hadrons, such as deuteron and $\rho$ meson,
one may also reach similar relations. Meanwhile, they provide some new structure functions which have no
analogue to the case of spin-1/2 targets~\cite{Hoodbhoy:1988am,Berger:2001zb,Cosyn:2017fbo}. \\

For a spin-1 target, there are 9 helicity nonflip GPDs and 9 helicity flip GPDs for each quark flavour (or for the gluon) at the twist-2 order. The spin-1 helicity nonflip (twist-2) GPDs are defined in Ref. \cite{Berger:2001zb} by considering the deeply virtual Compton scattering and meson electroproduction processes of the deuteron. Recently, the 9 helicity flip (twist-2) GPDs, or transversity GPDs, are introduced and discussed in Ref. \cite{Cosyn:2018rdm}. Among the total 9 helicity nonflip quark GPDs,
5 of them are unpolarized and 4 of them are polarized ones. The sum rules of the unpolarized GPDs can give the charge $G_C$,
magnetic $G_M$, and quadrupole $G_Q$ form factors. We have intensively studied those observables with a help
of a light-front constituent quark model for the $\rho$ meson phenomenologically~\cite{Sun:2017gtz}, where the
$\rho$ meson form factors $G_{C,M,Q}(Q^2)$, mean square charge radius, magnetic and quadrupole moments are
calculated. Our obtained results are reasonably compatible with the previous model calculations
and the experimental data~\cite{deMelo:1997hh,Gudino:2013jaa,Krutov:2018mbu}. Moreover, our calculated
results for the first Mellin moments of the unpolarized GPDs $H_1$ and $H_5$, which respectively correspond
to the reduced matrix elements and to the structure functions of $F_1$ and $b_1$ (the tensor structure),
are in a good agreement with the results from the Lattice QCD calculation~\cite{Best:1997qp}.
For the transversity GPDs of $\rho$ meson, they are remained to be studied. In this work, only helicity nonflip GPDs are considered. \\

To account for a polarized target, we know that the spin-dependent structure functions $g_1(x)$ and $g_2(x)$
are defined by the decomposition of the imaginary part of the forward virtual Compton scattering
amplitudes~\cite{Hoodbhoy:1988am,Jaffe:1989xx,Jaffe:1990qh,Guichon:1998xv}. In the leading order (twsit-2),
the forward limit of the polarized GPD $\tilde{H}_1(x,0,0)$ is related to $g_1(x)$
~\cite{Berger:2001zb,Best:1997qp}. It is believed that the $g_1$ gives the information of the polarized
quark density, namely, the probability to find a polarized quark (with longitudinal momentum fraction
$x$) parallel or antiparallel to the polarization of the target~\cite{Feynman:1973xc,Mankiewicz:1990ji}.
In addition, the sum $g_T=g_1+g_2$ involves the transverse spin density~\cite{Feynman:1973xc}.
In general, the structure functions, $g_2$, or $g_T$, also receive the contributions from a
quark-gluon correlation which comes from the twist-3 operator~\cite{Cortes:1991ja}. Thus, they may give
the information of the "high-twist effects" in a system. Many theoretical and experimental studies have
been preformed for both $g_1$ and $g_2$ (see for example
~Refs.~\cite{Song:1996ea,Dong:1997cdh,Anthony:1999py,Airapetian:2006vy,Adolph:2016myg})
in the literature. More details can be found in recent review
articles~\cite{Kuhn:2008sy,Chen:2010qc,Aidala:2012mv}. \\

To our knowledge, the spin-dependent structure functions $g_1$ and $g_2$ of spin-1 hadrons, particular
for the $\rho$ meson, have been rarely studied theoretically. Since we have successfully studied the unpolarized GPDs of the $\rho$ meson with a help of a light-front quark model, we extend our approach to
further calculate the polarized GPDs of the $\rho$ meson, and try to obtain its $g_1(x)$ from the forward
limit of the polarized GPDs $\tilde{H}_1(x,0,0)$. It is known that the spin structure function $g_2$ is
usually related to $g_1$ according to the Wandzura and Wilczek relation~\cite{Wandzura:1977qf}. However, as emphasized by Jaffe and Ji~\cite{Jaffe:1989xx,Jaffe:1990qh}, $g_2$ is not solely determined by $g_1$
as Wandzura and Wilczek concluded. There are another twist-2 function ($h_T$) and a twist-3 term which may
also have non-negligible contributions to $g_2$ (see Refs.
~\cite{Jaffe:1989xx,Jaffe:1990qh,Cortes:1991ja}). In this work, however, only twist-2 operators
are involved and we ignore $h_T$ and twist-3 terms as many other theoretical
calculations~\cite{Cortes:1991ja,Song:1996ea} did for simplicity. \\

In addition, the axial form factors for the spin-1 particle ${\tilde G}_{1,2}$ are seldom
discussed due to no axial current in electromagnetic interaction. However,
after taking into account the electro-weak interaction which contains axial vector currents, the two form
factors can be measured through the respond functions $W_{1,2,8}$~\cite{Pollock:1990uv}. This phenomenon is
similar to the nucleon (spin-1/2) case~\cite{Kaplan:1988ku}. Therefore, the axial form factors become
important when we study the electro-weak structure of the system, such as the parity violating in the
electron-deuteron scattering \cite{Ito:2003mr}. Since the axial form factors relate to the sum rules of
the polarized GPDs of the system, we may also estimate them according to our obtained polarized GPDs for
the $\rho$ meson. \\

This paper is organized as follows. In Section~\ref{sec:1}, the definitions and sum rules of the polarized
GPDs and the structure function $g_1$ etc. are briefly presented. Moreover, the light-front quark model
employed in this and our previous works is also shortly discussed in this section. In
Section~\ref{sec:evolution}, the evolution for the spin structure function $g_1$ is discussed.
Section~\ref{sec:Results} gives our numerical results for the polarized GPDs, the spin structure functions
$g_1$, $g_2$ and the axial form factors of the $\rho$ meson. Section~\ref{sec:conclusions} is devoted
for a short summary.\\

\section{Polarized GPDs and our model}
\label{sec:1}

\begin{figure}
\resizebox{0.4\textwidth}{!}{%
  \includegraphics{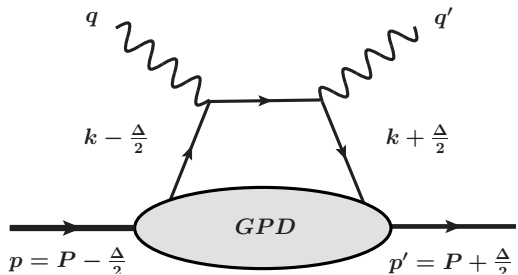}
}
\caption{The s-channel handbag diagram for GPDs. The u-channel one can be obtained by $q\leftrightarrow q'$.}
\label{fig:gpd}       % Give a unique label
\end{figure}

Fig. \ref{fig:gpd} illustrates the process we are considering. The notations are~\cite{Sun:2017gtz}
\eq
t &=&\Delta^2=(p'-p)^2=(q-q')^2 \ , \; \; Q^2=-q^2 \ ,  \nonumber \\
\xi &=&-\frac{\Delta\cdot n}{2P\cdot n}= -\frac{\Delta^+}{2P^+} \ , \ \ \abs{\xi}=\frac{\Delta^+}{2P^+} \ , \; \; (\,|\xi\,|\le1) \nonumber \\
x &=&\frac{k\cdot n}{P\cdot n}=\frac{k^+}{P^+} \ , \ \ \ \ \ \ (-1\le x\le1) \ ,
\en
where $p$ and $p'$ are the 4-momenta of the incoming and outgoing $\rho$ mesons, $P=({p'+p})/{2}$, $\Delta=p'-p
$, $n$ is a light-like 4-vector with $n^2=0$. Here $q$ is the virtual photon momentum, and $q'$ is treated as a real one.\\

The four polarized GPDs, for a spin-1 particle, are introduced in Ref.~\cite{Berger:2001zb},
\begin{eqnarray}
\label{eq:gpd}
\lefteqn{
\frac{1}{2} \int \frac{d \lambda}{2\pi}\, e^{ix (P z)}
  \langle p'|\, \bar{q}(-\half z)\, \slash{n} \gamma_5 \,
     q(\half z)\, \,|p \rangle \Big|_{z = \lambda n}
     } \nonumber \\
&=&- i \frac{\epsilon_{\mu \alpha \beta \gamma}
   n^\mu \epsilon'^{*\, \alpha}
  \epsilon^\beta P^\gamma}{P n}\,
\tilde{H}_1^q(x,\xi,t)
\nonumber \\
&&+ 2i\, \frac{\epsilon_{\mu \alpha \beta \gamma}\, n^\mu
 \Delta^\alpha P^\beta}{P n}\,
 \frac{ \epsilon^\gamma (\epsilon'^* P) +
        \epsilon'^{* \,\gamma} (\epsilon P) }{M^2}\,
\tilde{H}_2^q(x,\xi,t)
\nonumber \\
&&+ 2i\, \frac{\epsilon_{\mu \alpha \beta \gamma}\, n^\mu
 \Delta^\alpha P^\beta}{P n}\,
 \frac{ \epsilon^\gamma (\epsilon'^* P) -
        \epsilon'^{* \,\gamma} (\epsilon P) }{M^2}\,
\tilde{H}_3^q(x,\xi,t)
\nonumber \\
&&+ \frac{i}{2}\, \frac{\epsilon_{\mu \alpha \beta \gamma}\, n^\mu
 \Delta^\alpha P^\beta}{P n}\,
 \frac{ \epsilon^\gamma (\epsilon'^* n) +
        \epsilon'^{* \,\gamma} (\epsilon  n) }{P n}\,
\tilde{H}_4^q(x,\xi,t),
\end{eqnarray}
where $\epsilon_{0123}=1$ and $M$ is the $\rho$ meson mass. Without loss of generality, we choose the  $\rho^+$
meson in this work and omit the superscript hereafter when there is no ambiguity.  Thus, in the constituent
quark model, only $u$ and  ${\bar d}$ contribute to the current operator in Eq. (\ref{eq:gpd}).
Time reversal constraints that %$H_4$ and
$\tilde{H}_3^q$ are $\xi$-odd and all other GPDs $\xi$-even.  Taking the lowest moments of the polarized GPDs
in $x$, one recovers the axial vector form factors for each flavour $q$ \cite{Berger:2001zb},
\begin{eqnarray}
\label{eq:sum-rules}
\int_{-1}^1 dx\, \tilde{H}_i^q(x,\xi,t) = \tilde{G}_i^q(t) \ ,
\hspace{3em}     (i=1,2) ,
\end{eqnarray}
with matrix elements of
\begin{eqnarray}
\label{eq:axail-ff}
\lefteqn{
  \langle p' |\,
    \bar{q}(0)\, \gamma^\mu \gamma_5\, q(0) \,| p \rangle
= - 2i \, \epsilon^\mu{}_{\!\alpha \beta \gamma}\,
    \epsilon'^{* \alpha} \epsilon^\beta P^\gamma\; \tilde{G}_1^q(t)
}
\nonumber \\[0.2em]
&& {}+ 4i \, \epsilon^\mu{}_{\!\alpha \beta \gamma}\,
    \Delta^\alpha P^\beta\, \frac{\epsilon^\gamma (\epsilon'^*  P)
    + \epsilon'^{* \gamma} (\epsilon P)}{M^2}\; \tilde{G}_2^q(t) .
\end{eqnarray}

For other two GPDs, time reversal invariance gives
\begin{equation}
  \label{zero-sum-a}
\int_{-1}^1 dx\, \tilde{H}_3^q(x,\xi,t) = 0 \ ,
\end{equation}
and the Lorenz invariance constraints
\begin{equation}
  \label{zero-sum-b}
\int_{-1}^1 dx\, \tilde{H}_4^q(x,\xi,t) = 0 \ .
\end{equation}
\\

With respect to the axial-vector current $J^{5\mu}$,  one gets the axial vector form factors
\eq
\tilde{G}_{i} = \tilde{G}_{i}^u -\tilde{G}_{i}^d - \tilde{G}_{i}^s +\cdots \ , (i=1,2) \ ,
\en
where the definition for individual flavour is given in Eq.~(\ref{eq:axail-ff}). As shown later (in Eq.~(\ref{eq:axial_vector_iso_ff})), under the isospin symmetry,  $\tilde{G}_{i}^u =\tilde{G}_{i}^d$ in $\rho^+$ and the contributions
of light $u$ and $d$ quarks to the total axial vector form factors cancel each other.
When considering only the $u$ and $d$ flavours simultaneously, one gets
$\tilde{G}_{1,2}=0$~\cite{Pollock:1990uv}.

Due to the isospin symmetry and charge symmetry (G-parity), the polarized (or axial) GPDs are related by
\eq \label{eq:g_parity_axial}
\tilde{H}_{i,\rho^+}^u(x,\xi,t) = \tilde{H}_{i,\rho^+}^{d} (-x,\xi,t) \ ,
\en
where $i=1\sim4$.
Project the axial (polarized) GPDs onto isoscalar and isovector combinations, we have
\eq \label{eq:axial_isoscalar}
\tilde{H}_i^{I=0}(x,\xi,t)&=&\frac{1}{2} \left[ \tilde{H}_i^u(x,\xi,t) + \tilde{H}_i^{ d}(x,\xi,t) \right ] \ ,  \\
 \label{eq:axial_isovector}
\tilde{H}_i^{I=1}(x,\xi,t)&=&\frac{1}{2} \left[ \tilde{H}_i^u(x,\xi,t) - \tilde{H}_i^{ d}(x,\xi,t) \right ] \ ,
\en
and the corresponding axial vector isoscalar and isovector form factors are
\eq
\label{eq:axial_vector_isoscalar_ff}
\int_{-1}^1 dx\, \tilde{H}_i^{I=0}(x,\xi,t) &=& \tilde{G}_i^{u}(t) +  \tilde{G}_i^{d}(t) \equiv \tilde{G}_i^{I=0}(t)
\ , \\
\label{eq:axial_vector_isovector_ff}
\int_{-1}^1 dx\, \tilde{H}_i^{I=1}(x,\xi,t) &=& \tilde{G}_i^{u}(t) -  \tilde{G}_i^{d}(t) \equiv \tilde{G}_i^{I=1}(t) \ .
\en
With Eq. (\ref{eq:g_parity_axial}), one gets
\eq \label{eq:axial_isoscalar_1}
\tilde{H}_i^{I=0}(x,\xi,t) &=& \tilde{H}_i^{I=0}(-x,\xi,t) \ , \\
\label{eq:axial_isovector_1}
\tilde{H}_i^{I=1}(x,\xi,t) &=& -\tilde{H}_i^{I=1}(-x,\xi,t) \ ,
\en
which give
\eq \label{eq:axial_vector_iso_ff}
\tilde{G}_i^{I=0}(t) = 2\tilde{G}_i^u(t) \ , \ \tilde{G}_i^{I=1}(t) = 0 \ , \ (i=1,2) \ ,
\en
This results from $\tilde{G}_{i}^u =\tilde{G}_{i}^d$ in $\rho^+$.

For a comparison to the unpolarized case, we note that, for the unpolarized GPDs \cite{Sun:2017gtz}, there is
an overall minus sign difference w.r.t. Eq. (\ref{eq:g_parity_axial}) and Eq. (\ref{eq:axial_isovector}),
respectively,
\eq  \label{eq:g_parity_vector}
{H}_{i,\rho^+}^u(x,\xi,t) &=& -{H}_{i,\rho^+}^{d} (-x,\xi,t) \ ,\\
{H}_{i,\rho^+}^{I=1}(x,\xi,t) &=& {H}_{i,\rho^+}^{I=1}(-x,\xi,t) \ .
\en
where $i=1\sim5$. More details on the projection are referred to Refs.~\cite{Sun:2017gtz,Frederico:2009fk,Polyakov:1999gs}.
\\

As emphasized in Ref.~\cite{Pollock:1990uv}, the axial vector form factors $\tilde{G}_1$ and $\tilde{G}_2$ are
usually discarded  in the previous studies. After considering the electro-weak interaction,
one may expect nonzero strange quark contribution to $\tilde{G}_1$ and $\tilde{G}_2$, by
measuring the difference between the cross sections of the pure electromagnetic interaction and the
electro-weak interaction. These measurements can provide an important probe for the electro-weak structure
of the nucleons~\cite{Ito:2003mr}. For the $\rho$ meson, which is an isovector system, it is still quite
interesting to know what these two form factors, for $u$ and $d$ flavours, look like under our
phenomenological calculation.\\

In the forward limit $\Delta=0$, only $\tilde H_1^q$ survives and has quark density interpretation. Using the
relation of the helicity amplitudes for finding a quark in a $\rho$ meson \cite{Berger:2001zb}, one gets
\eq
\label{eq:ht1q}
{\tilde H}_1^q(x,0,0)= q^1_\uparrow(x) -q^1_\downarrow(x) \equiv \Delta q(x) \ ,
\en
where $x>0$ and $q^1_\uparrow(x)$ is the probability to find a quark with momentum fraction $x$ and polarization parallel to the $\rho$ meson helicity $+1$. Here $\Delta q(x)$ is called the spin dependent
density~\cite{Diehl:2003ny}, or the polarized quark distribution \cite{Ji:1998pc}. The parity constraints
$q^1_\uparrow=q^{-1}_\downarrow$. In the frame of GPDs, Eq. (\ref{eq:ht1q}) with $x~<~0$ stands for the
antiquark ($\bar q$) distribution at $-x$. This leads to the partonic decomposition
\cite{Diehl:2003ny,Ji:1998pc}
\eq \label{eq:delta_q}
{\tilde H}_1^q(x,0,0) = \theta(x) \Delta q(x) + \theta(-x) \Delta \bar q(-x) \ .
\en
By Eqs. (\ref{eq:g_parity_axial}) and (\ref{eq:delta_q}), one gets
\eq \label{eq:delta_u_d}
\Delta u_{\rho^+}(x)= \Delta{\bar d}_{\rho^+}(x) \ .
\en

As discussed in Ref.~\cite{Diehl:2003ny}, the $x$-even (``singlet'') combination
\eq
\tilde{H}_{1}^{q(+)}(x,\xi,t) = \tilde{H}_{1}^{q}(x,\xi,t) + \tilde{H}_{1}^{q}(-x,\xi,t)
\en
corresponds to the charge conjugation $C=+1$, and
gives $\tilde{H}_{1}^{q(+)}(x,0,0)=\Delta q(x)+\Delta \bar q(x)$ in the forward limit. The $x$-odd (``nonsinglet'' or ``valence'') combination
\eq
\tilde{H}_{1}^{q(-)}(x,\xi,t) = \tilde{H}_{1}^{q}(x,\xi,t) - \tilde{H}_{1}^{q}(-x,\xi,t)
\en
corresponds to the charge conjugation $C=-1$, and
gives $\tilde{H}_{1}^{q(-)}(x,0,0)=\Delta q(x)-\Delta \bar q(x)$ in the forward limit.
Thus, like the pion case \cite{Polyakov:1999gs,Broniowski:2007si}, for $\rho^+$, the valence (or nonsinglet) polarized quark distribution is
\eq \label{eq:valence}
\tilde V = \Delta u_{\rho} - \Delta \bar u_{\rho} + \Delta \bar d_{\rho} - \Delta d_{\rho} \ ,
\en
and the singlet polarized quark distribution is
\eq \label{eq:singlet}
\tilde S = \Delta u_{\rho} + \Delta \bar u_{\rho} + \Delta d_{\rho} + \Delta \bar d_{\rho} + \Delta s_{\rho} + \Delta \bar s_{\rho} \ .
\en
These two combinations do not mix under evolution (see Sec. \ref{sec:evolution}). The sea-quark distribution is defined as \cite{Broniowski:2007si}
\eq \label{eq:sea_quark}
\tilde s= \tilde S - \tilde V= 2(\Delta \bar u_{\rho} +  \Delta d_{\rho} ) + \Delta s_{\rho} + \Delta \bar s_{\rho} \ .
\en
In the present work, the $\rho^+$ meson is restricted to be only composed by an active quark $u$ and an active
antiquark $\bar d$, which means the contribution of sea quarks ($\bar u$, $d$, $s$ and $\bar s$) is not included
here.

On the other hand, at leading order, the polarized structure function $g_1^q (x)$ gives the fraction of spin carried by quarks~\cite{Best:1997qp}
\eq \label{eq:g1q}
g_1^q (x) = \frac{1}{2} \left[ q^1_\uparrow(x) -q^1_\downarrow(x)\right] + \{ q\rightarrow \bar{q} \} \ , %\nonumber  \\
%&=&\frac{1}{2} \Delta q(x) + \{ q\rightarrow \bar{q} \}  \ ,   \\
\en
and follows the relation~\cite{Berger:2001zb,Best:1997qp}
\eq
g_1(x)&=&\sum_q e_q^2 g_1^q (x) \ .
\en
Therefore, with Eqs.~(\ref{eq:ht1q}) and (\ref{eq:delta_u_d}),  we get
\eq
\label{eq:g1}
g_1(x) &=&
\frac{1}{2} e_u^2 \Delta u(x)+ \frac{1}{2} e_{\bar d}^2 \Delta {\bar d}(x) = \frac{1}{2} \left( e_u^2+ e_{\bar d}^2 \right)  \Delta u(x) \ , \\
%\nonumber \\
%&=& \frac{1}{2} \left( e_u^2+ e_{\bar d}^2 \right)  {\tilde H}_1^u(x,0,0)  \ , \; for~x~>~0 \ .
\label{eq:quark_spin}
\Delta q &\equiv& \int_0^1 \left[ g_1^u (x) +g_1^d (x) \right] dx= \int_0^1 \Delta u (x) \, dx\ .
\en
where $\Delta q $ is the total fraction of spin carried by valence $u$ and $\bar d$ in $\rho^+$.

In general, the rigorous expression for the structure function $g_2$ contains another twist-2 piece,
"transversity" $h_T$, and a twist-3 piece arising from quark-gluon correlation
~\cite{Cortes:1991ja,Song:1996ea}.  $h_T$ is proportional to the ratio of the current
quark mass to the target mass ($\sim m_c/M$) and it is commonly neglected in most studies \cite{Song:1996ea}.
In present work, both $h_T$ and the twist-3 parts are neglected, although it may not be small.
Under those approximations, one gets the Wandzura-Wilcze relation~\cite{Wandzura:1977qf} for $g_2$,
\eq
g_2^{WW}(x)=-g_1(x)+\int_x^1{{dy}\over y}g_1(y).
\en
Here, the $Q^2$-dependence is ignored, since at large $Q^2$, the $g_1$ and $g_2$ become scaling. It may not
be a good approximation to identify $g_2(x)=g_2^{WW}(x)$ (which may have $15\sim 40 \%$ breaking of the size of
$g_2$ \cite{Accardi:2009au}), however, we argue that it, at least, allows us to estimate the contribution of
the axial current operator to $g_2$. In this case, it is easy to verify the Burkhardt-Cottingham sum
rule~\cite{Burkhardt:1970ti} by changing the integral variables,
\eq
\int_0^1 g_2(x) \, dx=0 \ .
\label{eq:sumrule_g2}
\en
Notes that, according to Ref.~\cite{Jaffe:1989xx}, this relation remains to be tested since the derivation
in~\cite{Burkhardt:1970ti} is based on the assumption of the Regge theory. However, Ref. \cite{Song:1996ea} claims,
for proton, this sum rule for $g_2$  holds up to order of $O(M^2/Q^2)$. Finally, with those approximations, one gets the transverse spin density \cite{Feynman:1973xc,Wandzura:1977qf} 
$$
g_T(x) =g_1(x)+g_2(x) \sim \int_x^1{{dy}\over y}g_1(y).
$$

The Mellin moment of a function $f(x)$ is defined as
\eq
M_n(f)=\int_0^1 x^{n-1}f(x)dx \ .
\en
For the $\rho$ meson case, at the leading order (twist 2), one finds~\cite{Best:1997qp}
\eq
2M_n(g_1^q)=C_n^{(1)}r_n \  ,
\en
where $C_n^{(k)}=1+O(\alpha)$ is the Wilson coefficient of the operator product expansion and $r_n$ are the
reduced matrix elements.
These relations hold for both even and odd $n$-th orders with the quenched approximation.
Note that there are two different sets of notations labeling the moments of $F_1$, $b_1$ and $g_1$ respectively
in Refs.~\cite{Best:1997qp} and \cite{Mankiewicz:1990ji}. Here we follow the former. \\

\begin{figure}
\resizebox{0.5\textwidth}{!}{%
  \includegraphics{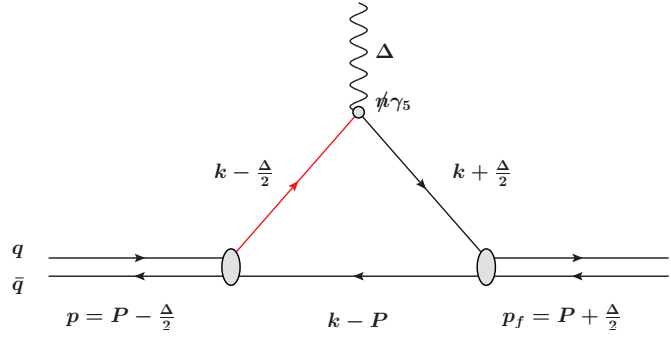}}
\caption{The struck $u$ quark in the valence regime for axial current. The momentum of the red line have positive plus component. }
\label{fig:gpd_u}       % Give a unique label
\end{figure}

In a numerical calculation, we employ the phenomenological light-front quark model to describe the interaction
between the spin-1 $\rho$ meson and its constitutes $u$ and ${d}$. It is based on a effective interaction
Lagrangian for the $\rho\rightarrow\bar{q}q$ vertex,
\begin{eqnarray}\label{key}
\lefteqn{
\mathcal{L}_I = -{\imath M\over f_\rho} \bar{q} \Gamma^\mu \mathbf{ \tau} q \cdot\mathbf{\rho}_\mu}
\nonumber\\ 
&&= -{\imath \sqrt{2}M\over f_\rho}
\left[ \frac{\bar{u}  \Gamma^\mu u
- \bar{d}  \Gamma^\mu d }{\sqrt{2}} \rho^0_\mu
+ \bar{u}  \Gamma^\mu d  \rho^+_\mu
+ \bar{d}  \Gamma^\mu u  \rho^-_\mu
\right] , 
\end{eqnarray}
where $\rho_\mu$ is the $\rho$ meson field, $f_\rho$ is the $\rho$ decay constant  (which may be absorbed in
the normalization factor $N$), and $\Gamma^\mu$ is a Bethe-Salpeter amplitude (BSA)
~\cite{Sun:2017gtz,Choi:2004ww},
 \begin{eqnarray}
 \Gamma^\mu=N\frac{
 \gamma^{\mu} - {(k_{q}+k_{\bar q})^{\mu}}/{(M_{i,f}+2m)}
 }{ [ k_{q}^2-m^2_R+ \imath \epsilon] [ k_{\bar q}^2-m^2_R+ \imath \epsilon] } \ ,
 \end{eqnarray}
where, for the $u$ quark contribution, the struck $u$ quark momentum $k_u=k-\Delta/2$ and the spectator constituent momentum is $k_s=k_{\bar{d}}=k-P$, as shown in Fig. \ref{fig:gpd_u}. $N$ is the normalization constant, $m$ and $m_R$ are the constituent quark and the regulator masses,
respectively, and $M_{i,f}$ are the kinematic invariant masses,~\cite{Sun:2017gtz,Choi:2004ww}
\begin{eqnarray}
\label{eq:vertexM:v}
M_{i}^2 = \frac{\kappa^2_\perp + m^2}{1-x'} + \frac{\kappa^2_\perp + m^2}{x'} \ , \\
M_{f}^2 = \frac{\kappa'^2_\perp + m^2}{1-x''} + \frac{\kappa'^2_\perp + m^2}{x''} \ ,
\end{eqnarray}
where the subscript $i(f)$ for initial(final) state and, following momenta convention in Fig. \ref{fig:gpd_u}, the
LF momentum fractions $x'(x'')$ and $\kappa_\perp(\kappa'_\perp)$ are
\begin{eqnarray} \label{eq:vertexM-variables}
x'&=& - {k_s^+\over p^+} ={1-x\over 1-\abs{\xi}}  \ , \; 
x''= x' {p^+\over p'^+} = {1-x\over 1+\abs{\xi}}  \ , \nonumber \\
\kappa_\perp &=& (k-P)_\perp- \frac{x'}{2} {\Delta}_{\perp} \ , \; 
\kappa'_\perp =%&=&
(k-P)_\perp+ \frac{x''}{2} {\Delta}_{\perp} \ . 
\end{eqnarray}
In the nonvalence regime where $-\abs{\xi}<x<\abs{\xi}$ leads to $x' > 1$ in Eq.~(\ref{eq:vertexM:v}) and (\ref{eq:vertexM-variables}), and the
initial vertex becomes the non-wave-function vertex.
To keep the mass square positive, as Refs.~\cite{Sun:2017gtz,Choi:2004ww}, we directly
replace $1-x'$ with $x'-1$ in Eq.~(\ref{eq:vertexM:v}) and get
\begin{eqnarray}
\label{eq:vertexM:nv}
M_{i(NV)}^2 = \frac{\kappa^2_\perp + m^2}{x'-1} + \frac{\kappa^2_\perp + m^2}{x'}.
\end{eqnarray}
Here, to keep this phenomenological $\Gamma^\mu$ respecting to the isospin symmetry (which is required
by  Eqs. (\ref{eq:g_parity_axial}), (\ref{eq:g_parity_vector}) and (\ref{eq:delta_u_d})), one has to employ
the symmetric momenta convention as shown in Fig. \ref{fig:gpd_u}. More details are explained in our
previous work~\cite{Sun:2017gtz}.

The expressions for individual axial GPDs can be obtained through the same way showed in the Appendix of Ref.~\cite{Sun:2017gtz}. For example, the $\tilde H_1^u$ is
\begin{eqnarray}\label{eq:u_quark}
\tilde H^{u}_1 (x, \xi, t) &=&
  N_{\mu\nu} \int \frac{d^4 k  }{ (2\pi)^4 }
  \, \delta \left[ n \cdot ( x P - k ) \right]
%  \nonumber \\  && \times
   Tr \Bigg[
  \frac{\imath ( \slash{k}-\slash{P}+m ) }{ (k-P)^2-m^2 + \imath \epsilon}
  \Gamma^\nu
%  \nonumber \\  && \times
  \frac{\imath ( \slash{k}+\frac{\slash{\Delta}}{2} +m ) }{ (k+\frac{\Delta}{2})^2 -m^2 + \imath \epsilon}
  \slash{n}\gamma_5
    \nonumber \\  && \times
  \frac{\imath ( \slash{k}-\frac{\slash{\Delta}}{2} +m ) }{ (k-\frac{\Delta}{2})^2 -m^2 + \imath \epsilon}
  \Gamma^\mu
  \Bigg],
%  \nonumber \\  %&&
\end{eqnarray}
where
\eq
N_{\mu\nu} &=& i \frac{M^2}{f_\rho^2}
  \frac{ c^2\, n^\alpha P^\beta  \epsilon_{ \alpha \beta \mu \nu}}
  { 4(2\pi)^3 \sqrt{\omega_{p'}\omega_{p} }\, (P \cdot n)}
  \ ,
\en
with $c$ being a normalization factor.

\section{On the QCD Evolution}
\label{sec:evolution}
Comparing the model-dependent results to the available "data", like the Lattice QCD calculation, one
may perform a QCD evolution to evolve the parton distribution and its moments from the factorization
scale $\mu_0$ to the scale that a Lattice QCD calculation is performed. For the calculated $\rho$ meson polarized
GPDs or structure functions in the present work, we compare our result with the Lattice QCD results at the
scale $\mu=2.4\gev$ with quenched approximation \cite{Best:1997qp}, as our previous work for the unpolarized
ones. Here, we ignore the gluon contribution to the evolution, thus, we can adopt the same (LO) DGLAP
evolution function for the moments of the single flavor structure function $g_1^u(x)$ as
\eq
\label{eq:dglap}
\frac{{\tilde V}_n^u(\mu)}{{\tilde V}_n^u(\mu_0)} = \left( \frac{\alpha(\mu)}{\alpha(\mu_0)}
\right)^{\gamma_n^{(0)}/(2\beta_0)} \ ,
\label{ratio}
\en
where the single quark spin fractions
$${\tilde V}_n^u=2 M_{n+1}\left[g_1^u(x)\right]
%=M_{n+1}\left[ {\tilde H}_{1}^u(x,0,0)\right]
\sim r_{n+1}
$$
and the running coupling constant is
\eq
\alpha(\mu) &=& \frac{4\pi}{\beta_0 \text{log}(\mu^2/\Lambda^2_{QCD})} \ , %\\
%\beta_0 &=& \frac{11}{3} N_c - \frac{2}{3} N_f \ ,
\en
where $\beta_0 = {11N_c}/{3}  - {2N_f}/{3} $ with $N_c=N_f=3$ and
\eq
\Lambda_{QCD} = 0.226~GeV
\en
being employed~\cite{Broniowski:2007si,Broniowski:2008hx}.
In our previous work, we performed the evolution of the Mellin moments of unpolarized structure function, and found the factorization scale of the model is
$%\eq \label{eq:scaleQ0}
\mu_0 = 528_{-62}^{+77} \; \text{MeV} \ .
$\\%\en

In our previous work, we obtained the evolution ratio for the valence quark distribution, by calculating the
evolution of the active $u$ quark unpolarized distribution. Here we adopt the same ratio for the evolution of
valence polarized quark distribution (or their Mellin moments) to compare with the Lattice QCD results
since the scale ($\mu=2.4\gev$) is same for both unpolarized and polarized cases.
In addition, the sea quark contributions (Eq.~(\ref{eq:sea_quark})) are excluded from our calculation, thus one
can observe that the nonsinglet (Eq. (\ref{eq:valence})) and singlet (Eq. (\ref{eq:singlet})) polarized quark
distributions make no more difference in present work.\\

\section{Numerical results}
\label{sec:Results}

\begin{figure*}
\centering
\subfigure[$ \xi=0 $]{\label{fig:ht1xi0}
\includegraphics[width=7.0cm]{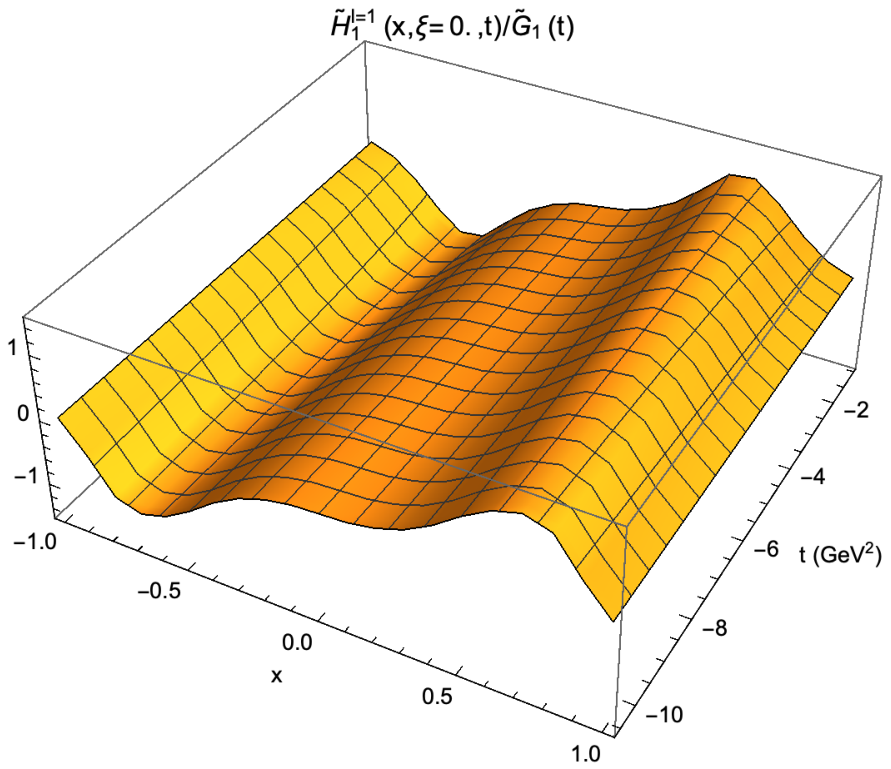}}
{\hskip 1cm}
\subfigure[$ \xi=-0.4 $]{\label{fig:ht1xi04}\includegraphics[width=7.0cm]{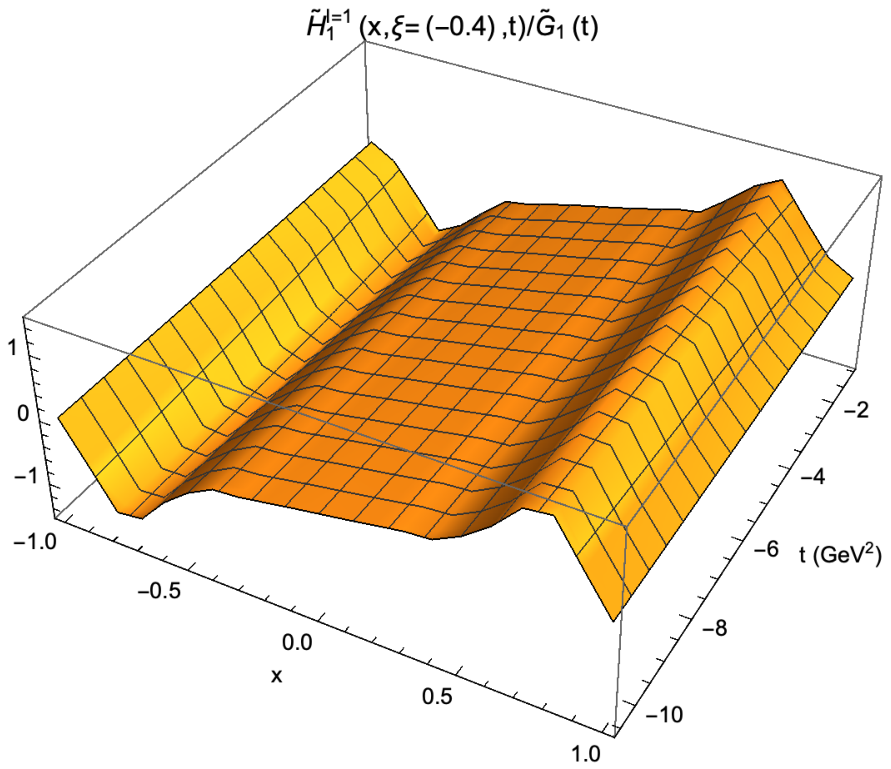}}
\caption{$\rho^+$ GPD ${\tilde H}_1$ with $\xi=0$ and $-0.4$. }
\label{fig:ht1}       % Give a unique label
\end{figure*}

\begin{figure*}
\centering
\subfigure[$ \xi=0 $]{\label{fig:ht2xi0}
\includegraphics[width=7.0cm]{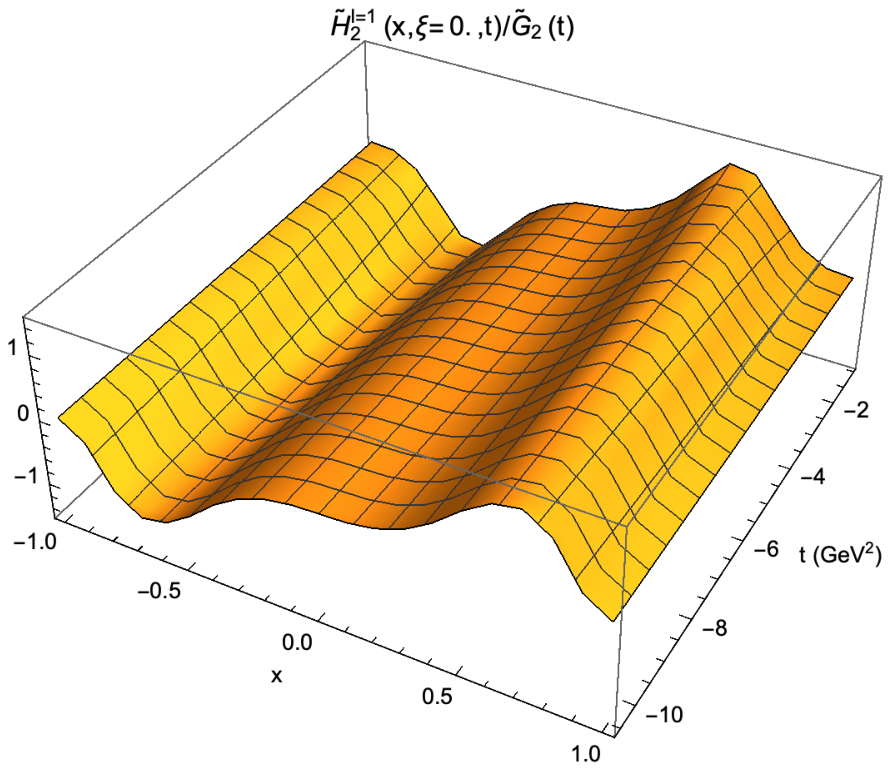}}
{\hskip 1cm}
\subfigure[$ \xi=-0.4 $]{\label{fig:ht2xi04}\includegraphics[width=7.0cm]{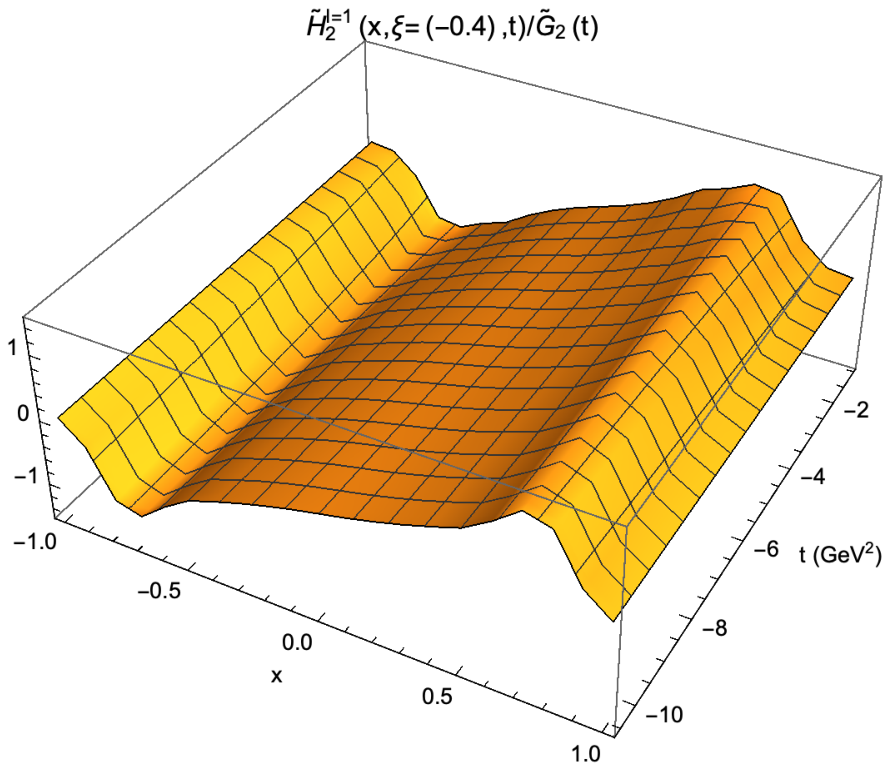}}
\caption{$\rho^+$ GPD ${\tilde H}_2$ with $\xi=0$ and $-0.4$. }
\label{fig:ht2}       % Give a unique label
\end{figure*}

\begin{figure}
\resizebox{0.5\textwidth}{!}{%
  \includegraphics{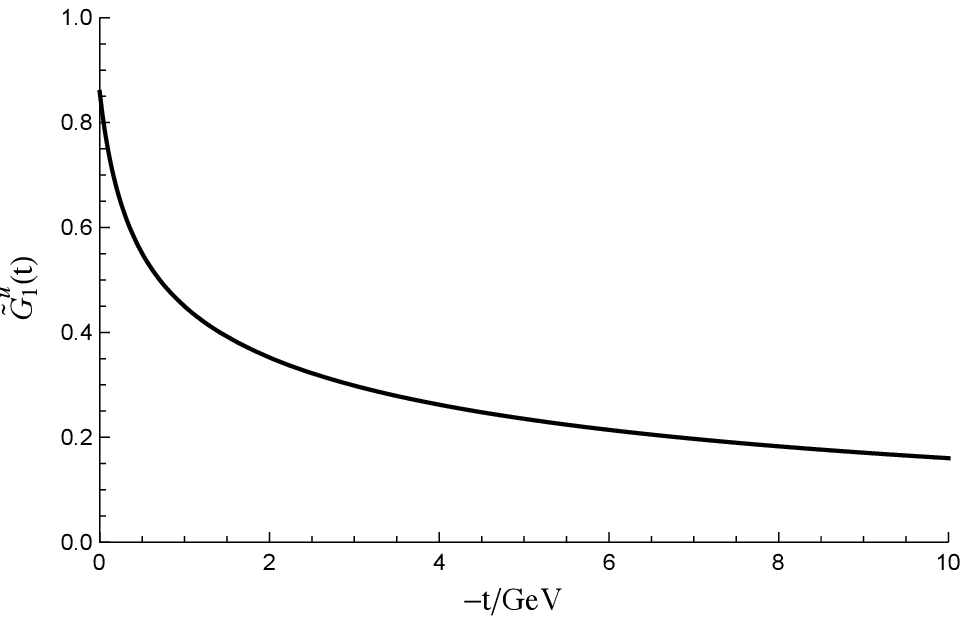}
}
\caption{The $u$ quark axial form factor ${\tilde G}_1^u(t)$. }
\label{fig:Gt1tu}       % Give a unique label
\end{figure}

\begin{figure}
\resizebox{0.5\textwidth}{!}{%
  \includegraphics{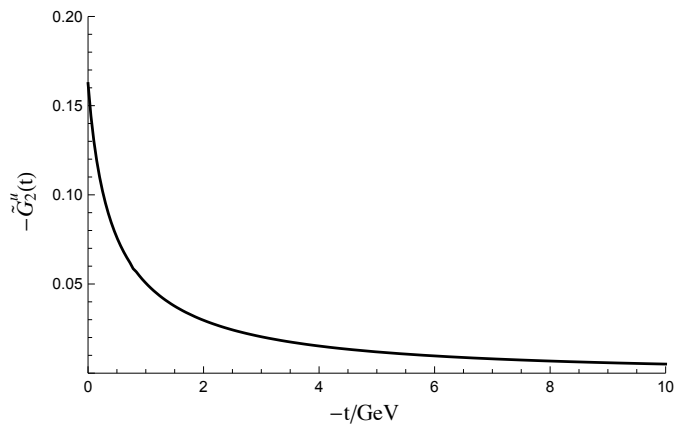}
}
\caption{The $u$ quark axial form factor ${\tilde G}_2^u(t)$}
\label{fig:Gt2tu}       % Give a unique label
\end{figure}

Following our previous work on the unpolarized GPDs~\cite{Sun:2017gtz}, we take the two model parameters, the
constituent mass $m=0.403~\gev$ and regulator mass $m_R=1.61~\gev$. We simply extend the model to the
polarized GPDs ${\tilde H}_{1,2}$ case. Their $x$- and $t$-dependences with skewness $\xi=0$ and $\xi=-0.4$ are
shown in Fig.~\ref{fig:ht1} and in Fig. \ref{fig:ht2} respectively. The results are normalized with respect
to the corresponding $u$ quark axial form factors. The obtained polarized GPDs have opposite values in the region $-1<x<0$ with respect to the region $0<x<1$ at the same $t$, as a consequence of the isospin symmetry of our model.
At the joint points of valance and non-valance regions, namely at $\abs{x}=\abs{\xi}=0.4$ in Figs.
\ref{fig:ht1xi04} and \ref{fig:ht2xi04}, our resulted ${\tilde H}_{1,2}$ are continuous.
This phenomenon fulfills the requirement of the consistency of the factorization at leading
twist~\cite{Diehl:2003ny}. Here, we take the momentum transfer $t$ up to $-10~\gev^2$, similar to the
unpolarized case. Comparing to the unpolarized GPDs, especially $H_1$, we find that the polarized GPDs
${\tilde H}_{1,2}$ vary much slow with respect to $t$. Figs.~\ref{fig:Gt1tu} and \ref{fig:Gt2tu} show the
single flavour axial form factor ${\tilde G}_1^u(t)$ and
${\tilde G}_2^u(t)$, respectively. Within the region $-10~\gev^2<t<0$, ${\tilde G}_1^u(t)$ is larger than
${\tilde G}_2^u(t)$ and decreases slower than ${\tilde G}_2^u(t)$ as $t$ increases. The starting points are
${\tilde G}_1^u(0)=0.86$ and ${\tilde G}_2^u(0)=-0.16$, respectively. Correspondingly, we have ${\tilde
G}_1^{I=0}(0)=1.72$ and ${\tilde G}_2^{I=0}(0)=-0.32$, respectively.\\

\begin{figure}
\resizebox{0.5\textwidth}{!}{%
  \includegraphics{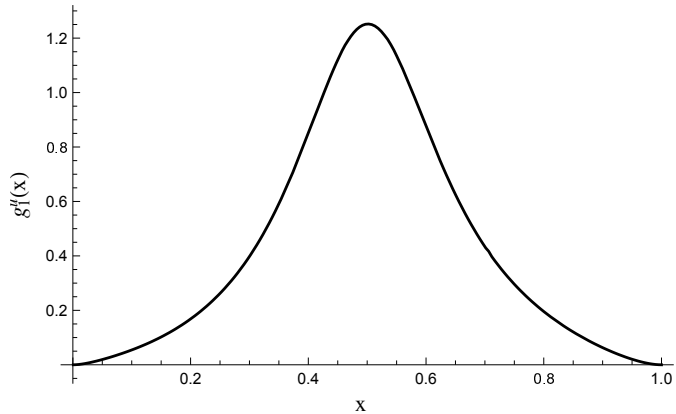}
}
\caption{The $u$ quark structure function $g_1^u(x)$}
\label{fig:g1}       % Give a unique label
\end{figure}

\begin{figure}
\resizebox{0.5\textwidth}{!}{%
  \includegraphics{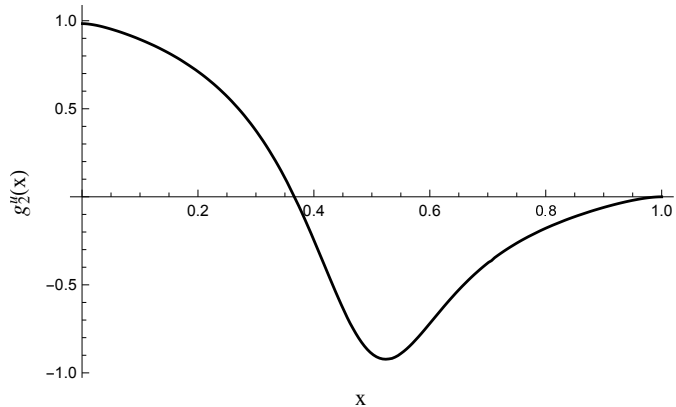}
}
\caption{The $u$ quark structure function $g_2^u(x)$}
\label{fig:g2}       % Give a unique label
\end{figure}

In Fig.~\ref{fig:g1} and Fig.~\ref{fig:g2}, the $x$ dependence of $g_1^u$ and $g_2^u$ are shown.
Our result for $g_1^u(x)$ remains positive in the whole $0<x<1$ region and it is nearly symmetry around $x=1/2$.
The available experimental data for deuteron $g_1^{(d)}(x)$, summarized in Ref. \cite{Airapetian:2006vy},
have negative values at small $x$ region, but it is believed to be consistent with zero after combining the
new COMPASS result \cite{Adolph:2016myg}. We think our result for the $\rho$ meson may indirectly confirm
the positiveness of $g_1(x)$.  In general, our twist-2 results for the $\rho$ meson $g_1$ have
similar $x$-dependence behavior with the $g_1$ of the deuteron in the new COMPASS result \cite{Adolph:2016myg}
(see its Fig. 4). Summing over $x$ of $g_1^u(x)+g_1^d(x)$ as Eq.~(\ref{eq:quark_spin}), we get
\eq
\Delta q= 0.86 \ .
\en
which means the fraction of spin carried by the constituent quark and antiquark in $\rho$ meson
is $0.86$, while the expected value is 1. This result is similar to the case of the nucleon (see for example
Ref.~\cite{Deur:2018roz}).
In general, the total fraction of spin carried by quarks and antiquarks in nucleon is not more than $30\%$
to $50\%$.  It is well known as the ``spin crisis" issue (or ``spin puzzle'')
~\cite{Ji:1998pc,Kuhn:2008sy,Aidala:2012mv,Deur:2018roz}.
As proposed by Sehgal \cite{Sehgal:1974rz}, another important contribution to the proton spin may come
from the orbital angular momentum of partons. Through the light-cone representation of the spin and orbital
angular momentum of relativistic composite systems, Brodsky, Hwang, Ma and Schmidt \cite{Brodsky:2000ii} have
shown that the ``spin crisis" of the nucleon can be explained due to the relativistic motion of quarks,
and the contribution of the orbital angular momentum. Thus the small  $\Delta q$ can be naturally understood.
According to Refs.~\cite{Schreiber:1988uw,Myhrer:2007cf}, the nucleon ``spin crisis"
maybe also be understood through the pion cloud effect together with relativistic corrections and one-gluon
exchange, which can transfer the quark spin to the orbital angular momentum and it mainly accounts for the
missing spin. The pions play a role of quark and antiquark sea.  Here, we suggest that the orbital
angular momentum may also be an important source for the $\rho$ meson spin and the corresponding parton splitting
processes $q \rightarrow qg$ and $g\rightarrow q\bar q$ responsible for the DGLAP evolution, generate the
orbital angular momentum \cite{Diehl:2003ny}.  After the evolution to a higher scale $\mu=2.4~\gev$,
as $r_1$ shown in Fig. (\ref{fig:rn}) later, $\Delta q$ becomes to around $60\%$.\\

Another way to understand the proton spin problem (see for example Refs.~\cite{Ma:1991xq,Ma:1992sj}) is to consider the Wigner rotation of the spin of a moving quark. In this sense, there is no need to require the
sum of quark's spin equals the total proton spin in the light front frame. \\

For the $g_2(x)$ structure function, the present constituent model predicts that
\eq  \int_0^1 g_2(x) \, dx=0.000112 \ ,
\en
comparing with the Burkhardt-Cottingham sum rule Eq. (\ref{eq:sumrule_g2}), we conclude that it is numerically
consistent with vanishing. With Eq. (\ref{eq:sumrule_g2}), we find that $g_2(x)$ has a remarkable feature of a
nontrivial zero between $x=0$ and $x=1$. Note again that $g_2$ should also receives contributions from
twist-3 quark-gluon correlation which may be not small comparing to that of the twist-2 piece.
The importance of this unique feature has stressed in previous works~\cite{Jaffe:1990qh,Cortes:1991ja,Anthony:1999py}.\\

\begin{figure}
\resizebox{0.5\textwidth}{!}{%
  \includegraphics{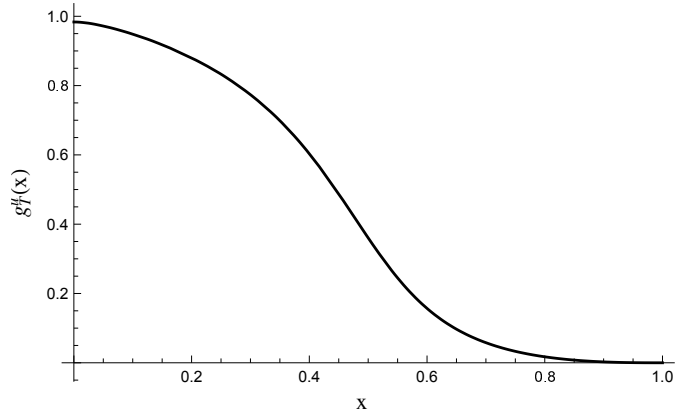}
}
\caption{The $u$ quark transverse spin density $g_T^u(x)$.}
\label{fig:gt}       
\end{figure}

\begin{figure}
\resizebox{0.5\textwidth}{!}{%
  \includegraphics{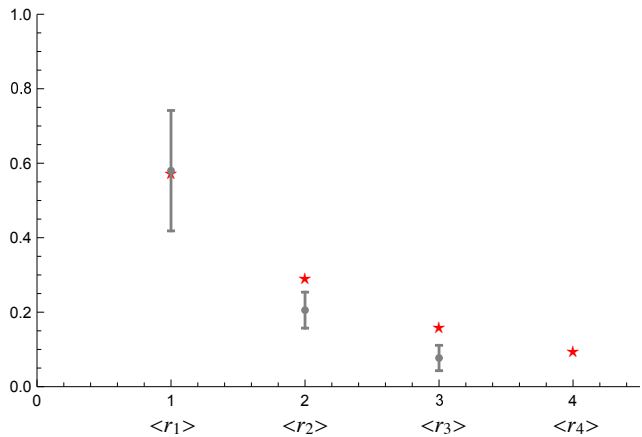}
}
\caption{$r_n$ for u quark. The red stars are our results and the gray ones with errors are
the Lattice QCD results \cite{Best:1997qp}.}
\label{fig:rn}       
\end{figure}

If one takes the massless limit of quark (asymptotic free), then $g_T=g_1+g_2$ would be small, but this
phenomenon contradicts to the $\rho$ meson rest mass, since the quarks are not free inside hadrons,
especially in the constituent quark model. Our results (see Fig. \ref{fig:gt}) tells that $g_T^u$ is sizeable in the
small and moderate $x$ regions ($<0.5$) and becomes much smaller in large $x$ region. It may be interpreted
that as the quark possesses more fraction of longitudinal momentum (larger $x$), it contributes less to the
transverse spin density. \\

The numerical evolution for the polarized structure functions is similar to the unpolarized case.
With the same ratio, which is 0.67, we evolute our results for the moments of $g_1$ to the scale of
the Lattice QCD result~\cite{Best:1997qp}. We compare the results of the two approaches in Fig.~\ref{fig:rn}.
The results of $r_n$ in Ref. \cite{Best:1997qp} was obtained with two sets of operators, and in
Fig.~\ref{fig:rn} we plot the averaged values. In general, our results agree with the Lattice
QCD ones. Moreover,  one more order of the moment (see $r_4$) is given by our calculation.

\section{Conclusions}
\label{sec:conclusions}

In this work, we extend our previous work on the $\rho$ meson GPDs with the light-front constituent quark model to
the polarized case. The polarized GPDs ${\tilde H}_{1,2}$ with nonzero skewness (e.g. $\xi=-0.4$) are given in
3-D plots w.r.t. $x$ and $t$. With the sum rules for ${\tilde H}_{1,2}$, we obtained the axial form factors
$\tilde G_{1,2}$, the spin structure functions $g_1(x)$ and $g_2(x)$, and the moments for
$g_1(x)$. After the evolution, our results of the moments of $g_1$ agree with the Lattice QCD results. The quark
spin contribution ($\Delta q=0.86$) to the $\rho$ meson spin and the transverse spin density $g_T$ for the
$\rho$ meson are also estimated with the constituent quark model for
the first time. The small value of $\Delta q$ for $\rho$ may be mainly explained by
its transfer to the orbital angular momentum carried by valence quarks, which is also a possible resolution of
the nucleon spin problem. Our numerical result for $g_2(x)$ shows that the Burkhardt-Cottingham sum rule holds
reasonably well in this work.

\section*{Acknowledgements}

One of the authors (BDS) acknowledges Ruhr-Universit\"at Bochum, for the warm hospitality where part of this work is being done; and thanks Bernard Pire and Maxim V. Polyakov
for helpful discussions.
This work is supported by the National Natural Science Foundation of China under Grant No. 11475192, by the fund provided to the Sino-German CRC 110 ``Symmetries and the Emergence of Structure in QCD" project by the NSFC under Grant No.11621131001,  and the Key Research Program of Frontier Sciences,  CAS, Grant No. Y7292610K1 and the State Scholarship Fund of China Scholarship Council No. 201804910428 and the DAAD Research Grants 2018 No. 57381332.

% Non-BibTeX users please use

\end{document}